\documentclass{aa}
\usepackage{graphicx,natbib}
\newcommand{\Msun}{\mbox{\rm $M_{\odot}$}}
\begin{document}
   \title{On high proper motion white dwarfs from photographic surveys}

   \author{C\'eline Reyl\'e \inst{1} \and Annie C. Robin \inst{1} 
	\and Michel Cr\'ez\'e \inst{2}}

   \institute{CNRS UMR6091, Observatoire de Besan\c{c}on, BP1615, 
    25010 Besan\c{c}on Cedex, France\\
    \and
    Universit\'e de Bretagne-Sud, BP 573, 
    F-56017 Vannes Cedex, France\\}
   \offprints{C\'eline Reyl\'e}

   \date{Received ; accepted }

   \titlerunning{On high proper motion white dwarfs}

   \abstract{The interpretation of high proper motion white dwarfs detected by 
Oppenheimer et al. (2001) was the start of a lively controversy. While 
the discoverers 
identify a large fraction of their findings as dark halo members, others 
interpret the same sample as essentially made of disc and/or thick disc stars.
We use the comprehensive description of Galactic stellar populations provided 
by the ``Besan\c{c}on'' model to produce a realistic simulation of Oppenheimer 
et al. data, including all observational selections and calibration biases. 
The conclusion is unambiguous: Thick disc white dwarfs resulting from ordinary 
hypotheses on the local density and kinematics are sufficient to explain the 
observed objects, there is no need for halo white dwarfs.  This conclusion is 
robust to reasonable changes in model ingredients.  The main cause of the 
misinterpretation seems to be that the velocity 
distribution of a proper motion selected star sample is severely biased in 
favour of high velocities. This has been neglected in previous analyses. 
Obviously this does not prove that no such objects 
like halo white dwarfs can exist, but Oppenheimer et al. observations drive 
their possible contribution in the dark matter halo down to an extremely low 
fraction.
   }

   \maketitle

\section{Introduction}

In a recent paper, \cite{Opp2001} (hereafter OHDHS) give evidence for a large 
number of faint white dwarfs detected in digitized photographic plates from
the SuperCOSMOS Sky Survey\footnote{see http://www-wfau.roe.ac.uk/sss}.
Their interpretation of these stars as dark halo members was the start of a 
controversy. While \cite{Reid2001} and more recently \cite{Graff2001} 
interpret OHDHS stars as the high velocity tail of the disc/thick 
disc populations, \cite{Koopmans2001} argue that there is a statistically 
significant break in the velocity distribution, which they interpret as the 
thick disc/halo break.
They propose a dynamical mechanism to produce these ``halo''
objects out of the disc. Another line of arguments has been developed by
\cite{Gibson2001} who point out a number of biases in OHDHS analysis 
all adding up to produce an overestimation of the local density of halo white 
dwarfs, but the authors do not discuss their nature. In addition, 
\cite{Hansen2001} finds a similar age distribution for this population as 
those in the thin disc white dwarf population.
In order to clarify the situation, we use the Besan\c{c}on model of stellar
populations in the Milky Way to simulate OHDHS observations.
In the next section we describe model hypothesis and the construction of
the simulated sample, the third section being dedicated to the discussion and
conclusions.

\section{Simulations}

We have performed simulations from the Besan\c{c}on Galactic model using the 
following assessment: disc stars are assumed to have ages from 0 to 10 Gyr,
with appropriate kinematics as a function of age 
\citep{Haywood1997A&A...320..440H}. The density laws are Einasto discs, which 
are close to sech$^2$ vertically \citep{Bienayme1987a,Bienayme1987b}. The 
white dwarf luminosity function is from \cite{Sion1977} and the 
photometry is derived from \cite{Bergeron1995} model atmospheres.
The ellipticities of these components are a function of age, and
constrained by the Galactic potential through the Boltzmann equation.
The average kinematics of white dwarfs in a volume limited sample in the solar
neighbourhood result in a velocity ellipsoid (42.1,27.2,17.2) and an 
asymmetric drift of 16.6 km s$^{-1}$.
 
The thick disc population is modeled as originating from a single epoch of
star formation. The adopted thick disc density law is described in 
\cite{Reyle2001}, as fitted to large sets of available star counts. 
The local density is 10$^{-3}$ pc$^{-3}$ or 7.1$\times 10^{-4}$ 
\Msun pc$^{-3}$ ($M_V<$8), that 
is about 6\% of the disc local density. In order to ensure continuity of the
density gradient at $z$~=~0, the density law is made quadratic near the plane
($z <$ 337 pc). Further out from the plane, the density law follows an
exponential of scale height 850 pc.

The white dwarf luminosity function (hereafter WDLF) of the thick disc 
has been computed by 
Garcia-Berro (private communication) from the model of \cite{Isern1998} 
assuming a Salpeter initial mass function (IMF) and an age of 12 
Gyr. We attempted to normalise the WDLF by computing the number of 
white dwarfs 
relative to main sequence stars, assuming that all stars with a mass greater
than the mass at the turnoff ($M_{\mbox{\scriptsize{TO}}}$ = 0.83 \Msun) are now white dwarfs.
We consider a two slope IMF, the lower mass parts being constrained
by available star counts \citep{Reyle2001} accounting for a binary correction:

$dN/dm \propto  m^{-2.35}; M_{\mbox{\scriptsize{TO}}} < m < 8 \Msun$

$dN/dm \propto  m^{-0.75}; 0.1 \Msun < m <   M_{\mbox{\scriptsize{TO}}}$

The ratio of white dwarfs to main sequence stars is therefore:\\

$\frac{N_{\mbox{\scriptsize{WD}}}}{N_{\mbox{\scriptsize{SP}}}} = \frac{0.74 
\int_{M_{\mbox{\scriptsize{TO}}}}^{8M_{\odot}} m^{-2.35} dm}
{\int_{0.1M_{\odot}}^{M_{\mbox{\scriptsize{TO}}}} m^{-0.75} dm}$ = 0.43

The factor 0.74 ensures the continuity at the turnoff. 
However, it appears that the predicted number of thick disc white 
dwarfs, with these assumptions, is much higher than the number of 
observed white dwarfs in the OHDHS sample. Several 
reasons could be invoked: First, we do not know how the OHDHS sample is 
complete, specially near the limiting magnitude. Second, hypotheses which have 
led to the estimation of the thick disc WDLF have 
not been tested yet because these stars are not identified as such in current 
white dwarfs samples: Uncertainties remain on the IMF slope 
(at high mass, with
no observational constraints, we have chosen the conservative Salpeter IMF, 
while at
low mass an IMF slope $\alpha$~=~1, still within our error bars, drives
down the fraction to 0.33). Uncertainties remain also
on the star formation history, which may be more complex than a single burst.
These reasons led us to normalize the thick disc WDLF in order to be 
in agreement with the OHDHS sample: the fraction of 
white dwarfs over sub-turnoff stars drops down to 20\%, about the value
of \cite{Hansen2001}, corresponding to a local density 
of 5$\times 10^{-4}$ stars pc$^{-3}$. This normalisation is also 
in agreement with
the number of white dwarfs in the expanded LHS (Luyten Half Second proper 
motion catalogue) white dwarf sample 
\citep{Liebert1999}.
The velocity ellipsoid of the thick disc is constrained by photo-astrometric 
survey in several Galactic directions \citep{Ojha1996} to be (67,51,42) while 
the asymmetric drift is 53 km s$^{-1}$.

Dark halo white dwarfs are modeled assuming that the dark halo is filled with 
ancient white dwarfs with a certain factor $f$. The dark halo local density is 
8~$\times~10^{-3}$~\Msun~pc$^{-3}$ from dynamical constraints 
\citep{Bienayme1987a}. Hence the local density
of halo white dwarfs is assumed to be $f$ times this value. As a first guess
compatible with microlensing constraints, we set $f$~=~0.1 in this
simulation. We use \cite{Bergeron1995} model atmospheres for hydrogen white 
dwarfs of mass 0.6 \Msun to the turnover ($M_V <$ 17), and
\cite{Chabrier2000} models for the cooler part of the sequence. 
The halo white dwarfs 
luminosity function has been computed by \cite{Isern1998}
with the \textit{ad hoc} \cite{Chabrier1999} IMF and an 
age of 12 or 14 Gyr. The space velocities of the dark halo are 
computed assuming the same kinematics
as for the spheroid: A velocity ellipsoid (131,106,85) and a null
rotation, the LSR rotational velocity being 229 km s$^{-1}$
\citep{Bienayme1987a}.

The colours are computed by transforming the standard 
Johnson-Cousins system to the observed one using the equations given in
the SuperCOSMOS Sky Survey description:

$B_J = B - 0.28 \times (B-V); \mbox{for} -0.1<B-V<1.6$

$R59F = R-0.006-0.059\times(R-I)+0.112\times(R-I)^2$\\
$+0.0238\times(R-I)^3$

The selection of the simulated sample has been done strictly following the
OHDHS process. We simulate a field of 4165 deg$^2$ at the 
south Galactic cap. We select a sample of stars with R59F $\leq$ 19.45, and 
proper motions in the range 0.33 to $3''$~yr$^{-1}$, assuming a mean error 
of $0.04''$~yr$^{-1}$. 
In order to reproduce the space velocities deduced by OHDHS, from the 
resulting sample, we compute the reduced proper motions, absolute magnitudes
$M_{B_J}$, distances and space velocities U 
and V using their formulae. 

\section{Results and discussion}

\begin{figure*}
\centering
\includegraphics[width=17cm]{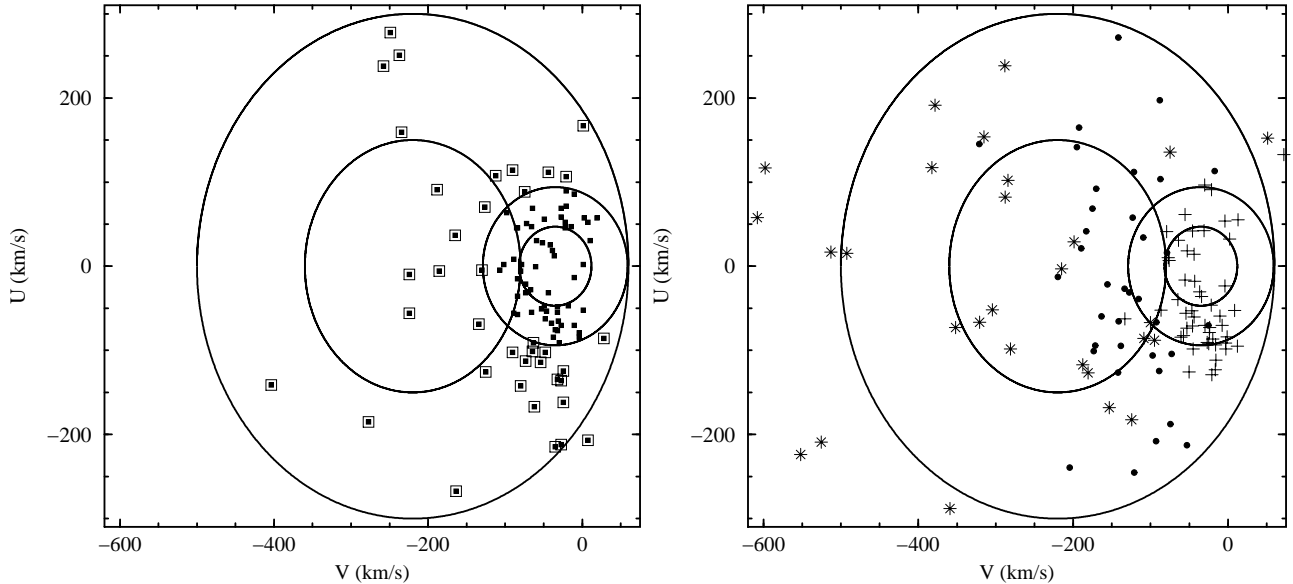}
   \caption{Velocity space distribution of white dwarfs in the Oppenheimer et
al. (2001) sample: a. Original observations. Objects with boxes are 
interpreted as halo stars. b. Simulation with a white dwarf 
local density being 10\% ($f$~=~0.1) of the dark halo local density, and an age
of 12 Gyr. Crosses: disc stars, circles: thick disc stars, asterisks: dark 
halo stars.}
   \label{fig1}
\end{figure*}

Fig.\ref{fig1}a reproduces the observed distribution of the OHDHS sample
 in the (U,V) plane. The ellipses indicate the 1$\sigma$ and 2$\sigma$ 
velocity 
dispersions as defined by OHDHS for the old disc (right) and the halo (left). 
Objects out of the 2$\sigma$ old disc ellipse are interpreted as dark halo 
white dwarfs by OHDHS (squares with boxes). 
This figure can be compared with the simulated distribution of the selected 
white dwarfs represented in fig.\ref{fig1}b. The disc is coded with crosses, 
the thick disc with circles, and the dark halo with asterisks. 
Ellipses are the same as in fig.\ref{fig1}a for comparison.

The vast majority of OHDHS white dwarfs out of the 2$\sigma$ old disc ellipse
can in fact be interpreted as thick disc white dwarfs, as expected from 
ordinary hypotheses on the local density and kinematics of the thick disc 
white dwarfs, although it is not possible to be sure that none of these white 
dwarfs belong to the dark halo. 
\cite{Hansen2001} argues that the OHDHS white dwarfs are not old enough to 
account for the thick disc population, given that most of the white dwarfs
in the thick disc should be fainter in the case of a single burst of star 
formation in the thick disc 12~Gyr ago. However, the simulation shows that,
even in the burst case, the number of younger white dwarfs in the thick disc
remains large enough to explain the OHDHS sample. They come from stars for
which the time spent on the main sequence is not negligible. Actually, very 
few of the simulated stars are beyond the turnover in the white dwarf 
colour-magnitude diagram.

Another difference that appears in fig.\ref{fig1} is the number of
high velocity stars:
Considering an age of 12 Gyr for the halo white dwarfs and a fraction 
$f$~=~0.1, 10 objects are expected to have V~$<~-$400~km~s$^{-1}$ (9
simulations were performed to compute a mean value for this number of high 
velocity objects).
None of the OHDHS white
dwarfs are present in this part of the diagram, indicating that the 
halo white dwarf local density is no more than 1\% of the dark halo local 
density, in agreement with the values derived by 
\cite{Koopmans2001} and \cite{Gibson2001}. 
However, this upper-limit remains uncertain as the 
completeness of the OHDHS sample is not precisely determined. Furthermore, 
if considering an halo age of 14~Gyr, the expected number is 2.

The photometric estimation of the distances could introduce a bias. They are determined 
from a linear colour-magnitude relation, with uncertainties around 20\%.
The linear relation is no longer valid for old white dwarfs beyond the
turnover in the colour-magnitude diagram. Using this relation leads to an 
overestimation of their luminosities and consequently, their distances and V 
velocities. However, very few stars on the OHDHS sample are affected by this 
bias (see figure 1 of \cite{Koopmans2001}). On the contrary, the simulated
sample considering a 12 Gyr dark halo with $f$~=~0.1 contains some of these 
old white dwarfs, which are the stars one is looking for to explain the dark 
matter halo.

Another strong bias comes from the selection of high proper motion stars which
isolates a high velocity tail of each population. The ellipses in 
fig.\ref{fig1}
characterize kinematics of a complete halo/thick disc sample, but are 
incorrect for a
high proper motion sample. 
The simulated sample can be separated in three distinct populations 
and it well reproduces the observed dispersions and rotation lag. Stars within 
the 2$\sigma$ old disc ellipse are mainly disc stars, in 
contradiction with \cite{Koopmans2001} who neglected this population. Out of
the 2$\sigma$ old disc ellipse, stars are dominated by thick disc stars and a
few disc stars. The fraction $f$~=~0.1 for an halo of 12~Gyr clearly 
overestimates the number of observed stars in the 2$\sigma$ halo ellipse and 
even more outside.

As shown in table~1, due to the observational biases in the sample, the mean 
asymmetric drift of the resulting sample is much larger that the normal value 
for a complete sample in the solar neighbourhood. 
The expected values for halo white dwarfs are also given in table~1. There are
much larger than the values found by OHDHS, 
implying that the halo population cannot dominate the sample.

It should also be noted that, even if a large uncertainty remains 
on the true local density of thick disc white dwarfs, our conclusions
are robust to changes of hypothesis within this uncertainty, since they are
mainly based on the velocity characteristics of 
the sample. Uncertainties on the thick disc kinematics also exist.
Adopting a larger asymmetric drift, such as 89 km~s$^{-1}$ (as in 
\cite{Robin1993}), 
would slightly displace the thick disc white dwarfs towards negative V, 
but our conclusions remain unchanged. Had we adopted an
asymmetric drift small enough to displace most of the thick disc stars inside
the 2$\sigma$ old disc ellipse, the fraction $f$=0.1 would 
underestimate the number of stars with 
$-400$~km~s$^{-1}<$V$<-200$~km~s$^{-1}$, but a higher value for 
$f$ would dramatically overestimates the number of high velocity stars with 
V$<-400$~km~s$^{-1}$.

\begin{table}
\label{table1}
\caption{Rotational lag and velocity dispersion $\sigma_U$ for disc,
thick disc and halo white dwarfs. Values in a complete sample are those
in the original model. Values from the selected sample come from the simulation
with the high proper motion and apparent magnitude selections. In the selected
sample the computed velocity dispersion and lag are about twice what it would 
be in a complete sample.
}
\begin{tabular}{cccc}
\hline
		& 	    &complete sample &selected sample\\
\hline
	 	&disc	    &16.6 km s$^{-1}$ &30 km s$^{-1}$  \\
lag	        &thick disc &53 km s$^{-1}$   &122 km s$^{-1}$ \\
		&halo       &229 km s$^{-1}$  &336 km s$^{-1}$ \\
\hline
		&disc	    &42.1 km s$^{-1}$ &64 km s$^{-1}$  \\
$\sigma_U$	&thick disc &67 km s$^{-1}$   &112 km s$^{-1}$ \\
		&halo       &131 km s$^{-1}$  &207 km s$^{-1}$ \\
\hline
\end{tabular}
\end{table}

\section{Conclusions}

The simulated sample, taking into account all observational bias, gives an 
unambiguous result: Most high velocity
white dwarfs in the OHDHS sample can be safely interpreted as 
disc and thick disc stars. The expected contribution of 
spheroid white dwarfs, assuming their mass density to be 1.3$\times 10^{-5}$ 
\Msun pc$^{-3}$ \citep{Gould1998} would be 0.5 stars. The OHDHS sample 
has been misinterpreted because the bias introduced by the selection of high 
proper motions has been neglected.
Although some of the stars in the sample may be part of the halo, it is not 
necessary to call for exotic objects such as white dwarfs in the dark halo.
However, this sample provides a direct observation of thick disc white dwarfs. 
Such observations will lead to the construction of the luminosity function
for thick disc white dwarfs, and hence to a better understanding of the Milky 
Way's history.

The simulation also shows that a sample dominated by the dark halo
white dwarfs would have much larger
velocity dispersions than the OHDHS sample. The number of expected candidates
with a V velocity less than $-400$~km~s$^{-1}$, compared with the null detection, 
implies that the fraction of dark matter halo made of such objects is less 
than 1\% for a 12 Gyr halo, or that the halo is older than 12 Gyr.

In a more general way, this study shows that care must be taken when one tries
to interpret the nature of high proper motion white dwarfs in photographic 
surveys (e.g. \cite{Knox1999,Ibata2000,Monet2000}), as the thick disc 
population is a non-negligible part of high proper motion selected sample,
and as thick disc white dwarfs cannot be easily distinguished from halo white 
dwarfs. As also shown by \cite{Creze2001}, 
the search for 
halo white dwarfs needs to be performed in deeper astrometric surveys, when 
the thick disc becomes non dominant over the expected halo white dwarfs 
population in high proper motion selected samples.

\begin{acknowledgements}
The authors thank G. Chabrier and L. Koopmans for fruitful discussions.
\end{acknowledgements}

\end{document}